\documentclass[runningheads]{llncs}
\usepackage{graphicx}
\graphicspath{ {.} }
\usepackage{amsmath}
\usepackage{bookmark}
\usepackage{times}
\usepackage{xcolor}
\usepackage{verbatim}
\usepackage{xcolor}
\usepackage{soul}
\usepackage[utf8]{inputenc}
\usepackage[small]{caption}
\usepackage{amssymb}
\usepackage{csquotes}
\usepackage{fancyvrb}
\usepackage{listings}
\usepackage{svg}
\usepackage{tikz}
\usepackage{algpseudocode}
\usepackage{calc}
\usepackage{newfloat}
\usepackage{marvosym}
\DeclareFloatingEnvironment[placement={!ht},name=Listing, within=none]{instructionListing}
\DeclareMathOperator*{\argmax}{arg\,max} 
\newcommand{\R}{\mathbb{R}}

\begin{document}
%
\title{Implementing Argumentation-enabled Empathic Agents}
 
 \author{Timotheus Kampik \Letter \orcidID{0000-0002-6458-2252} \and
     Juan Carlos Nieves\orcidID{0000-0003-4072-8795} \and
     Helena Lindgren\orcidID{0000-0002-8430-4241}}
     \authorrunning{T. Kampik et al.}
     \institute{Umeå University, 901 87, Umeå, Sweden \\
     \email{\{tkampik,jcnieves,helena\}@cs.umu.se}}
\authorrunning{T. Kampik et al.}

\maketitle              
\begin{abstract} 
    In a previous publication, we introduced the core concepts of \emph{empathic agents} as agents that use a combination of utility-based and rule-based approaches to resolve conflicts when interacting with other agents in their environment.
    In this work, we implement proof-of-concept prototypes of empathic agents with the multi-agent systems development framework \emph{Jason} and apply argumentation theory to extend the previously introduced concepts to account for inconsistencies between the beliefs of different agents. We then analyze the feasibility of different admissible set-based argumentation semantics to resolve these inconsistencies.
    As a result of the analysis, we identify the \emph{maximal ideal extension} as the most feasible argumentation semantics for the problem in focus.

    \keywords{Agent architectures \and Agent-oriented software engineering \and Argumentation}
\end{abstract}
\section{Introduction}
Complex information systems that act with high degrees of autonomy and automation are ubiquitous in human society and shape day-to-day life.
This leads to societal challenges of increasing frequency and impact, in particular when systems are primarily optimized towards simple metrics like views and clicks and are then used by malevolent actors for deceptive purposes, for example, to manipulate political opinions. 
Emerging research highlights these challenges and suggests the development of new multi-agent system concepts to address the problem (\cite{sen_agents_2018}, \cite{kampik_coercion_2018}).
In a recent publication, we outlined the basic concepts of empathic agents \cite{kampik_towards_2018}, based on an established definition of empathy as the ability to \textquote{simulate another's situated psychological states, while maintaining clear self–other differentiation} \cite{doi:10.1111/j.2041-6962.2011.00056.x}\footnote{We based our empathic agent on a rationality-oriented definition of empathy, to avoid the technical ambiguity definitions that focus on emotional empathy imply. A comprehensive discussion of definitions of empathy is beyond the scope of this work.}.
Our empathic agent is utility-based but additionally uses acceptability rules to avoid conflicts with agents that act in its environment.
Thereby, we attempt to emulate empathic human behavior, which considers rules and societal norms, as well as the goals and intentions of others.
We consider our agent a research contribution to the \textquote{synergistic combination of modelling methods} for \textquote{autonomous agents modelling other agents} as outlined as an open research problem in the discussion of a survey by Albrecht and Stone \cite{albrecht_autonomous_2018}.

This paper addresses the following research questions:
\begin{itemize}
    \item How can proof-of-concept prototypes of empathic agents be implemented with a multi-agent systems development framework?
    \item How can abstract argumentation (see: Dung \cite{dung_acceptability_1995}) be used to extend the agents to be able to resolve scenarios in which the beliefs of different agents are initially inconsistent?
    \item Which admissible set-based argumentation semantics are most feasible for resolving these inconsistencies?
\end{itemize}

To illustrate our work in a context that can be considered of societal relevance, we introduce the following example scenario:
A \emph{persuader} agent can select exactly one item from a list of persuasive messages (\emph{ads}) it can display to a \emph{mitigator} agent.
Based on the impact the message will have on a particular end-user the mitigator represents, the mitigator will either accept the message and allow the end-user to consume the persuader's service offering (and the message), or terminate its session with the persuader.
The persuader's messages are considered \emph{empathic agent actions}; the mitigator does not \emph{act} in the sense of the empathic agent core framework (see: Section~\ref{concepts}).
If the mitigator accepts the action proposal of the persuader, the environment will pay different utility rewards (or punishments) to persuader and mitigator.
The goal of the persuader is to select a message that is utility-optimal for itself, considering that the mitigator has to accept it.
We suggest that the scenario, albeit simple, reflects a potential type of real-world use case for future empathic agents of greater maturity.
Improving the end-user experience of systems that are traditionally primarily optimized towards one simple metric can be considered a societally beneficial use case for empathic agents:
(self-)regulation could motivate advertisement-financed application providers to either implement both persuader and mitigator to create a more purposeful user experience, or to open up their APIs to third-party clients that try to defend the interests of the end-users.

The rest of this paper is organized as follows:
First, we elaborate on the core concepts of the empathic agent (Section~\ref{concepts}), taking into consideration the belief-desire-intention (BDI) architecture in which the concepts are to be implemented.
Then, we describe the implementation of the concepts with the multi-agent systems development framework Jason~\cite{bordini2005bdi} (Section~\ref{jason}).
Next, we apply argumentation theory and extend the core concepts and their implementation with Jason to handle inconsistencies between the beliefs of different agents (Section~\ref{argumentation}).
To clarify our choice of argumentation semantics, we then analyze the applicability of different admissible set-based semantics (Section~\ref{argumentation-analysis}).
Subsequently, we examine other work related to argumentation with Jason, briefly discuss the work in the context of existing negotiation approaches, discuss limitations of the implemented agents, and outline possible steps towards more powerful empathic agents (Section~\ref{discussion}), before we conclude the paper (Section~\ref{concl}).

\section{Empathic Agent Concepts}
\label{concepts}
In this section, we elaborate on the empathic agent core concepts as initially sketched out in \cite{kampik_towards_2018}.

\noindent A set of interacting empathic agents can be described as follows:
\begin{itemize}
    \item   In an environment, $n$ empathic agents $\{A_0, ..., A_n\}$ are acting.
    \item   Each agent $A_i ( 0 \leq i \leq n )$ can execute a finite set of actions \\ $Acts_{A_i} = \{Act_{A_{i_0}}, ..., Act_{A_{i_m}}\}$. All agents execute all actions simultaneously at one specific point in time; (inter)actions in continuous time or over a series of discrete time steps are beyond the scope of the basic concept.
    \item   In its \emph{belief base}, each agents has a utility function that assigns a numeric value to all possible action combinations: $u_{A_i}: Acts_{A_0} \times ... \times Acts_{A_n} \rightarrow \{-\infty, \R, +\infty\}$.
    \item   Each agent strives to determine a set of actions it should execute that is considered \emph{acceptable} by all agents.
            For this, an agent $A_i$ determines the sets of actions that maximize its own utility function ($\argmax u_{A_i}$)\footnote{Note that $\argmax u_{A_i}$ returns a set of sets.}:
            If there is no own best set of actions $acts_{am_{i,k}} \in \argmax u_{A_i}$ that matches a set of best actions of all of the other agents, a conflict of interests exists.
            Hence, a conflict of interest can be determined by the following boolean function:
            \begin{align*}
                &c (\argmax u_{A_0}, ..., \argmax u_{A_i}, ..., \argmax u_{A_n}) = \\
                & \quad true, if : \\
                &  \quad \quad \not \exists \ acts_{am_{0,l}}, ..., acts_{am_{i,k}}, ..., acts_{am_{n,m}} \\
                &  \quad  \quad  \quad in \argmax u_{A_0} \times ... \times \argmax u_{A_i} \times ... \times \argmax u_{A_n}: \\
                &  \quad  \quad  \quad \{ acts_{am_{0,l}} \cap ... \cap acts_{am_{i,k}} \cap ... \cap acts_{am_{n,m}} \} \ne \{ \} \\
                & \quad else: false
            \end{align*}
    \item   In addition to its utility mapping, each agent $A_i$ has a set of acceptability rules $Accs_{A_i}$ in its belief base.
            An acceptability rule $Acc_{A_{i_j}} \in Accs_{A_i}$ is a boolean function that takes as it input the to-be-executed actions and returns a boolean value\footnote{Note that a single acceptability rule does not necessarily consider all to-be-executed actions, i.e. it might ignore some of its input arguments.}:
            $Acts_{A_0} \times ... \times Acts_{A_n} \rightarrow \{true, false\}$.
    \item   If the conflict determination function does not determine a conflict, the agents can go ahead and execute their optimal actions.
    If the conflict determination function returns $true$ (determines a conflict), each agent applies the acceptability rules to check whether the actions it would ideally execute are in fact not acceptable or whether they can \emph{potentially} be executed nevertheless.
    Here, the agents can employ different game-theoretical approaches (for example: minimizing the maximal loss) for ensuring that the execution of conflicting, but acceptable action sets by multiple agents does not lead to utility outcomes that are \textquote{bad} (in particular: worse than executing the actions that maximize combined utility) for all parties\footnote{As the simple examples we implement in this paper feature only one \emph{acting} agent (a second agent is merely approving or disapproving of the actions), such game-theoretical considerations are beyond scope.
    Hence, we will not elaborate further on them.}.
    If the actions are not acceptable or are acceptable but chosen to not be executed, each agent can either choose to maximize the shared utility of all agents ($\argmax (u_{A_0} \times ... \times u_{A_n})$) or to select the next best action set and execute the conflict and acceptability check for them.
    We refer to agent variants that employ the former approach as \emph{lazy} empathic agents, and agents that use the latter as \emph{full} empathic agents.
    Lazy empathic agents save computation time by not iterating through action sets that do not optimize the ideally possible individual utility but provide better individual utility than the action sets that maximize combined utility. However, they might not find \textquote{good} solutions that are represented by action sets they do not cover\footnote{For now, we assume all agents in a given scenario have the same implementation variant. Empathic agents that are capable to effectively interact with empathic agents of other implementation variants or with non-empathic agents are--although interesting--beyond scope.}.
\end{itemize}
Listing~\ref{alg:1} describes the empathic agent base algorithm (\emph{lazy} variant) in a two-agent scenario.
\begin{instructionListing}
    \caption{Empathic agent base algorithm}
    \label{alg:1}
    \begin{enumerate}
        \item Determine actions that maximize own utility ($\argmax u_{A \: self}$).
        \item Check for conflict with actions that maximize utility of other agent ($\argmax u_{A \: other}$).
        \item If conflicts appear (apply conflict determination function $c$): \\
        \phantom{-----} Check if actions that maximize own utility are acceptable despite conflicts. \\
        \phantom{-----} if conflicts are acceptable \\
        \phantom{-----} (apply acceptability rules as function $Acts_{self} \times Acts_{other} \rightarrow \{true, false\}$): \\
        \phantom{----------} Execute $(\argmax u_{A \: self}) \cap Acts_{self}$. \\
        \phantom{-----} else: Execute $(\argmax (u_{A \:self} \times u_{A \: other})) \cap Acts_{self}$. \\
        else: Execute $(\argmax u_{A \: self}) \cap Acts_{self}$ (or $(\argmax (u_{A \:self} \times u_{A \: other})) \cap Acts_{self}$).
    \end{enumerate}
\end{instructionListing}

\section{Implementation of an Empathic Agent with Jason}
\label{jason}
To provide proof-of-concept prototypes of empathic agents, we implemented a simple empathic agent example scenario with Jason, a development framework that is well-established in the multi-agent systems community\footnote{The implementation of our empathic agents with Jason (including the Jason extension we introduce below, as well as a technical report that documents the implementation) is available at \url{https://github.com/TimKam/empathic-jason}.}.

\begin{figure}
    \fontsize{7}{10}\selectfont
    \centering 
    \def\svgscale{0.6}
    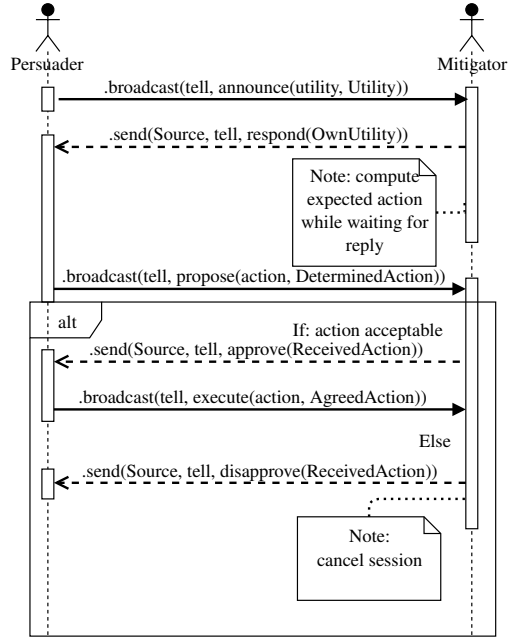
    \caption{Empathic agent interaction: sequence diagrams, basic example}
    \label{sequence_basic}
\end{figure}

The agents implement the scenario type we explained in the introduction as follows:
\begin{enumerate}
    \item
     The persuader agent has a set of utility mappings--(\emph{utility value}, \emph{unique action name})-tuples--in its belief base\footnote{The mappings are end-user specific. In a scenario with multiple end-users, the persuader would have one set of mappings per user.}:

    \begin{Verbatim}[fontsize=\small]
        revenue(3, "Show vodka ad").
        revenue(1, "Show university ad").
        ...
    \end{Verbatim}

    The mappings above specify that the persuader can potentially receive three utility units for showing a vodka advertisement, or one utility unit for showing a university advertisement.
    The persuader communicates these utility mappings to the mitigator.
    \item
    The mitigator has its own utility mappings:

    \begin{Verbatim}[fontsize=\small]
        benefit(-100, "Show vodka ad").
        benefit(10, "Show university ad").
        ...
    \end{Verbatim}

    The mitigator's utility mappings are labelled \emph{benefit}, in contrast to the persuader's \emph{revenue} label.
    While this should reflect the difference in impact the actions have on persuader and mitigator (or rather: on the end-user the mitigator is proxying), it is important to note that we consider the utility of the two different mappings \emph{comparable}.
    The mitigator responds to the persuader's announcement by sending back its own mapping.
    In addition to the utility mappings, both persuader and mitigator have a set of \emph{acceptability rules} in their belief base\footnote{Note that in Jason terminology, acceptability rules are \emph{beliefs} and not \emph{rules}.}
    :

    \begin{Verbatim}[fontsize=\small]
        acceptable(
            "Show university ad",
            "Show community college ad").
        acceptable(
            "Show community college ad",
            "Show university ad").
        ...
    \end{Verbatim}
    
    The \emph{intended meaning} the empathic agents infer from these rules is that both agents agree that \emph{Show university ad} is always acceptable if the preferred action of the mitigator is \emph{Show community college ad} and vice versa.
    After sending the response, the mitigator determines the action it thinks should be executed (the \emph{expected} action), using the algorithm as described in Section \ref{concepts}.
    In the current implementation, the agents are \emph{lazy} empathic agents, i.e. they will only consider the actions with maximal individual and maximum shared utility (as explained above)
    \footnote{If at any step of the decision process, several actions could be picked because they provide the same utility, the agents will always pick the first one in the corresponding list to reach a deterministic result.}.
    \item The persuader determines its action proposal in the same way and announces it to the mitigator.
    \item When the response is received and the own expected action is determined, the mitigator compares the received action with the determined action.
    If the actions are identical, the mitigator sends its approval of the action proposal to the persuader.
    If the actions are inconsistent, the mitigator sends a disapproval message and cancels the session.
    \item If the persuader receives the approval from the mitigator, it executes the agreed-upon action.
\end{enumerate}
Figure \ref{sequence_basic} shows a sequence diagram of the basic empathic agent example we implemented with Jason.
The message labels show the internal actions the agents use for communicating with each other.

\section{Argumentation for Empathic Agents: Reaching Consensus in Case of Inconsistent Beliefs}
\label{argumentation}
In real-world scenarios, agents will often have inconsistent beliefs because of an only partially observable environment or subjective interpretations of environment information.
Then, further communication is necessary for the agents to reach consensus on what actions to execute and to prevent mismatches between the different agents' decision making processes that lead to poor results for all involved agents.
In this section, we show how abstract argumentation can be used to synchronize inconsistent agents \emph{acceptability rules} in the agents' belief bases.
I.e., we improve the \emph{autonomy} for the agent from the perspective of moral philosophy in that we enhance ability of an empathic agent to \textquote{impose the [...] moral law on [itself]}~\cite{sep-autonomy-moral}.

Abstract argumentation, as initially introduced by Dung formalizes a theory on the exchange of arguments and the acceptability of arguments within a given context (\emph{argumentation framework}) as a tuple $\langle A, R \rangle$ where $A$ is a finite set of arguments (for example: $\{a, b, c\}$ and $R$ denotes the arguments' binary attack relations (for example: $\{(a, b), (b, c)\}$, i.e. \emph{a attacks b and b attacks c}). Dung introduces different properties of argument sets in argumentation frameworks, such as \emph{acceptability}, \emph{admissibility}, and \emph{completeness}~\cite{dung_acceptability_1995}.
Dung's initial theory has since been extended, for example by Bench-Capon, who introduced value-based argumentation to account for values and value preferences in argumentation~\cite{bench2003persuasion}.
In multi-agent systems research, argumentation is a frequently applied concept, \textquote{both for automating individual agent reasoning, as well as multiagent interaction}~\cite{Rahwan:2013}.
It can be assumed that applying (abstract) argumentation to enhance the consensus-finding capabilities of the empathic agents is an approach worth investigating.

\begin{figure}
    \centering
    \fontsize{7}{10}\selectfont
    \centering 
    \def\svgscale{0.6}
    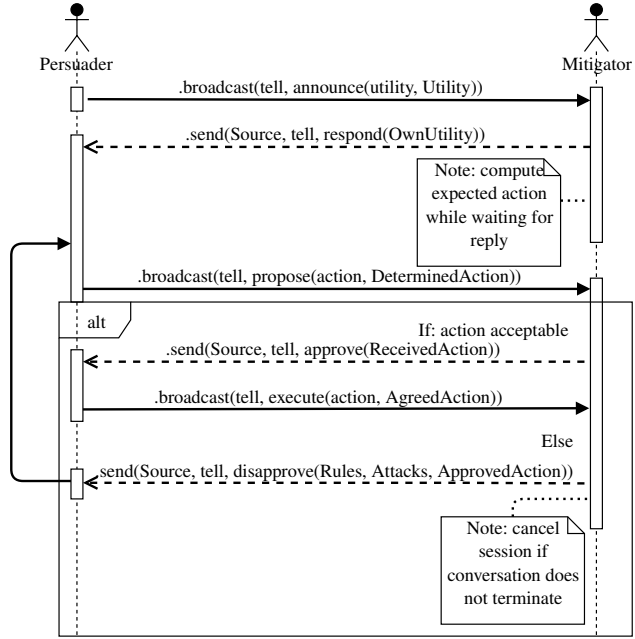
    \caption{Empathic agent interaction: sequence diagrams, argumentation example}
    \label{sequence_argumentation}
\end{figure}

In the previous example implementation, which primarily implements the empathic agent core concepts, inconsistencies between acceptability rules are irreconcilable (and would lead to the mitigator's rejection of the persuader's action proposal).
To address this limitation, we use an abstract argumentation approach based on Dung's argumentation framework \cite{dung_acceptability_1995} and its \emph{maximal ideal extension} definition \cite{dung2007computing}.
We enable our agents to launch attacks at acceptability rules.
An attack is a tuple $(a, b)$, with $a$ being the identifier of an action or of another attack and $b$ being the identifier of the attack itself.
Each attack is added to a list of arguments, with the rule it attacks as its target.
Attacked rules are added as arguments without attack targets.
Additional arguments can be used to launch attacks on attacks.
The argumentation framework is resolved by determining its \emph{maximal ideal extension}\footnote{Note that we compare different argumentation semantics in Section~\ref{argumentation-analysis}.}.
The instructions in Listing~\ref{alg:2} describe the algorithm the argumentation-enabled empathic agents use for belief synchronization. The algorithm is executed if the agents initially cannot agree on a set of actions, which would lead to a cancellation of the interaction using only the basic approach as introduced above.
\begin{instructionListing}
\caption{Argumentation-enabled belief synchronization}
\label{alg:2}
\begin{enumerate}
    \item \textbf{Mitigator agent} (After having received not acceptable action proposal): Request acceptability rules from mitigator.
    \item \textbf{Persuader agent}: Send acceptability rules.
    \item \textbf{Mitigator agent} (Upon receiving acceptability rules):
        \begin{enumerate}
            \item Consider acceptability rules as initial arguments $\{acc_0, ..., acc_n\}$ in argumentation framework $AF = \langle \{acc_0, ..., acc_n\}, \{\} \rangle$.
            Resolve $AF$ by determining maximal ideal extension.
            \item Decide whether: \\
            \phantom{-----} new attacks should and can be launched, \\
            \phantom{-----} or $AF$ can be accepted as-is, \\
            \phantom{-----} or session should be cancelled.
            \item \label{go-to-m} If attacks should be launched: \\
            \phantom{-----} Add additional set of arguments to $AF$ that attack acceptability rules that are \\
            \phantom{-----} inconsistent with own beliefs. Send updated $AF$ to persuader (go to~\ref{go-to-p}). \\
                    else if $AF$ can be accepted as-is: \\
                    \phantom{-----} Accept and resolve $AF$. \\
                    else: \\
                    \phantom{-----} Cancel session.
        \end{enumerate}
    \item \textbf{Persuader agent} (Upon receiving $AF$):
    \begin{enumerate}
        \item \label{go-to-p} Generate/update own argumentation framework $AF$. Resolve $AF$ by determining maximal ideal extension.
        \item Decide whether: \\
            \phantom{-----} new attacks should and can be launched, \\
            \phantom{-----} or $AF$ can be accepted as-is, \\
            \phantom{-----} or session should be cancelled.
        \item If attacks should be launched: \\
        \phantom{-----} Add additional set of arguments to $AF$ that attack acceptability rules that are \\
        \phantom{-----} inconsistent with own beliefs. Send updated $AF$ to mitigator (go to~\ref{go-to-m}). \\
                else if $AF$ can be accepted as-is: \\
                \phantom{-----} Accept and resolve $AF$. \\
                else: \\
                \phantom{-----} Cancel session.
    \end{enumerate}

\end{enumerate}
\end{instructionListing}
We extended Jason to implement an argumentation function as an internal action.
The function calls the argumentation framework solver of the \emph{Tweety Libraries for Logical Aspects} \emph{of Artificial Intelligence and Knowledge Representation} \cite{thimm_tweety_2014}, which we wrapped into a web service with a RESTful HTTP interface.
In our example implementation, we extend our agents as follows:
\begin{enumerate}
    \item The persuader starts with some acceptability rules in its belief base that are not existent in the belief set of the mitigator:
    
    \begin{Verbatim}[fontsize=\small]
        acceptable("Show vodka ad", "Show university ad").
        ...
    \end{Verbatim}

    In the provided example, the persuader believes (for unknown reasons) that the action \emph{Show vodka ad} is acceptable even if the action \emph{Show university ad} provides greater utility to the mitigator.
    \item When the mitigator assesses the action proposal, it detects that its own expected action (\emph{Show university ad}) is inconsistent with the persuader's expected action (\emph{Show vodka ad}).
    It sends a disapproval message to the persuader that includes an attack on the acceptability rules of the \emph{Show vodka ad} action:

    \begin{Verbatim}[fontsize=\small]
        attack("Show vodka ad", "Alcoholic").
    \end{Verbatim}

    \item The persuader then constructs an argumentation framework from acceptability rules and attacks.
    With the help of the Jason argumentation extension, it determines all preferred extensions of the argumentation framework and removes the arguments (acceptability rules and attacks) that are not part of any preferred extension.
    Subsequently, the persuader updates its belief base accordingly, re-determines the action proposal, and sends the proposal to the mitigator, who then re-evaluates it (and ideally accepts it).
    Potentially, the persuader could also launch attacks on the mitigator's attacks and so on, until consensus is reached or the session is aborted\footnote{However, the provided example code implements only one argumentation cycle.}.
\end{enumerate}
Figure \ref{sequence_argumentation} shows a sequence diagram of the argumentation-capable empathic agent example we implemented with Jason.
\section{Argumentation Semantics Analysis}
\label{argumentation-analysis}
In this section, we explain why our argumentation-enabled empathic agents use argumentation semantics that select the \emph{maximal ideal} extension.
The purpose of the required argumentation reasoner is to remove all arguments from an argumentation framework that have been attacked by arguments that in turn have not been successfully attacked.
I.e., the naive requirements are as follows:
\begin{enumerate}
    \item Given a set of arguments $S$, the semantics must exclude all successfully attacked arguments from $S$, and only these.
    \item An argument in $S$ is considered successfully attacked if it is attacked by any argument in $S$ that is not successfully attacked itself.
\end{enumerate}
According to Dung, a \textquote{preferred extension of an argumentation framework AF is a maximal (with respect to set inclusion) admissible set of AF}.
In this context, \textquote{a conflict-free set of arguments S is admissible iff each argument in S is acceptable with respect to S}, given an \textquote{argument $A \in AR$ is acceptable with respect to a set S of arguments iff for each argument $B \in AR$: if B attacks A then B is attacked by S}~\cite{dung_acceptability_1995}.
As Dung's definition of preferred extensions is congruent with the stipulated requirements, it seems reasonable for the empathic agent implementations to resolve argumentation frameworks by using the corresponding argumentation semantics.

However, if there are arguments that attack each other directly or that form any other circular structure with an even number of nodes, multiple preferred extensions exist.
For example, given the arguments $\{a, b\}$ and the attacks $\{(a, b), (b, a)\}$ both $\{a\}$ and $\{b\}$ are preferred extensions.

This behavior implies ambiguity, which prevents the empathic agents from reaching a deterministic resolution of their belief inconsistency.
As this is to be avoided, we introduce an additional requirement:
\begin{enumerate}
    \setcounter{enumi}{2}
    \item The semantics must determine exactly one set of arguments by excluding ambiguous arguments.
\end{enumerate}
Dung defines the grounded extension of an argumentation framework as \textquote{the least (with respect to set inclusion) complete extension} and further stipulates that an \textquote{admissible set S of arguments is called a complete extension iff each argument, which is acceptable with respect to S, belongs to S}~\cite{dung_acceptability_1995}.
In the above example, the grounded extension is $\{\}$.
As determining the grounded extension avoids ambiguity, using grounded semantics seems to be a reasonable approach at first glance.
However, as Dung et al. show in a later work, grounded semantics are \emph{sceptical} beyond the stipulated requirements, in that they exclude some arguments that are not successfully attacked~\cite{dung2007computing}\footnote{In Example~\hyperref[subsubsection:example2]{2}, we illustrate an empathic agent argumentation scenario, in which grounded semantics are overly strict.}.
In the same work, the authors define a set of arguments $X$ as \emph{ideal} \textquote{iff X
is admissible and it is contained in every preferred set of arguments} and show that the maximal ideal set is \textquote{a proper subset of all preferred sets}~\cite{dung2007computing}.
Hence, we adopt maximal ideal semantics for determining the set of acceptable arguments of our empathic agents.

The following two running examples highlight the relevance of the distinction between preferred, maximal ideal, and grounded semantics in the context of our empathic agents.

\subsubsection*{Example 1}
\label{subsubsection:example1}
\begin{figure}
    \begin{tikzpicture}[
        roundnode/.style={circle, draw=black!60, minimum size=15mm},
        allnode/.style={dashed, circle, fill=lightgray, draw=black!60, minimum size=15mm, font=\bfseries},
        prefnode1/.style={circle, draw=black!60, minimum size=15mm, font=\bfseries},
        prefnode2/.style={circle, fill=lightgray, draw=black!60, minimum size=15mm}
        ]
    \node[prefnode2]    (a1)    at(0,0)  {a1};
    \node[prefnode1]    (b1)    at(0,3)  {b1};
    \node[prefnode2]    (c1)    at(3,3)  {c1};
    \path [->]  (b1) edge node[left] {} (a1);
    \path [->]  (c1) edge node[left] {} (b1);
    \path [->]  (b1) edge node[left] {} (c1);

\end{tikzpicture}
\caption{Example 1: argumentation semantics visualization: preferred (bold and grey, respectively). Grounded and maximal ideal extensions are $\{\}$. The preferred argument sets are also complete. \{\} is an additional complete set of arguments.}
\label{fig:example1}
\end{figure}
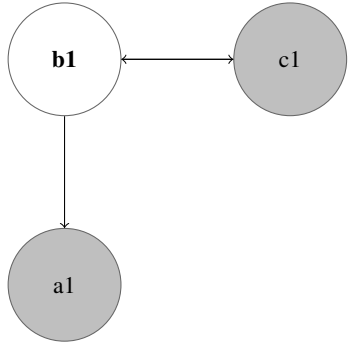

Example 1 highlights the advantage of grounded and maximal ideal semantics over preferred and complete semantics for the use case in focus. The to-be-solved empathic agent scenario is as follows:
\begin{itemize}
    \item Acceptability rules persuader: $\{(Show \: steak \: ad, *)\}$ ($a_1$)\footnote{For the sake of simplicity, we use a wild card ($*$) to denote that the acceptability rule applies no matter which preference the mitigator agent has. Note that this syntax is not supported by our implementation.}
    \item Acceptability rules mitigator: $\{\}$.
\end{itemize}
To resolve the acceptability rule inconsistency, the agents exchange the following arguments:
\begin{enumerate}
    \item Mitigator: $\{(a_1, Steak \: ad \: too \: unhealthy)\}$ ($b_1$)
    \item Persuader: $\{(b_1, Steak \: ad \: healthy \: enough)\}$ ($c_1$)
    \item Mitigator: $\{(c_1, b_1)\}$ 
\end{enumerate}
As a result, we construct the following argumentation framework:
\begin{align*}
    AF_1 = \langle \{a_1, b_1, c_1 \}, \{(b_1, a_1) (c_1, b_1), (b_1, c_1) \} \rangle
\end{align*}
Applying preferred, complete, grounded, and maximal ideal semantics to solve the framework yields the following results:
\begin{itemize}
    \item Preferred: $\{b_1\}, \{a_1, c_1\}$
    \item Complete:  $\{b_1\}, \{a_1, c_1\}, \{\}$
    \item Grounded: $\{\}$
    \item Maximal ideal: $\{\}$
\end{itemize}
As can be seen, applying preferred or complete semantics does not allow for a clear decision on whether the acceptability rule should be discarded or not, whereas both grounded and maximal ideal semantics are addressing this problem by discarding all arguments that are ambiguous in complete or ideal semantics, respectively.
Figure~\ref{fig:example1} contains a graphical overview of how the different argumentation semantics solve the argumentation framework of the first example.

\subsubsection*{Example 2}
\label{subsubsection:example2}
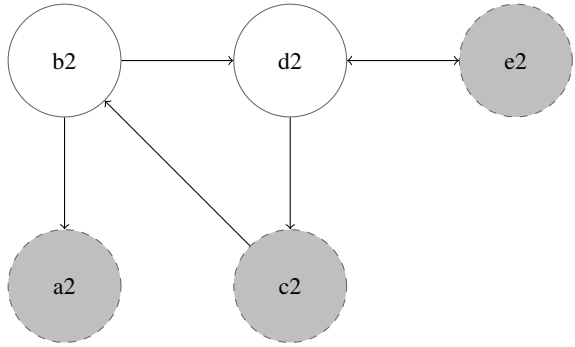
\begin{figure}
\begin{tikzpicture}[
    roundnode/.style={circle, draw=black!60, minimum size=15mm},
    allnode/.style={dashed, circle, fill=lightgray, draw=black!60, minimum size=15mm},
    prefnode2/.style={circle, fill=lightgray, draw=black!60, minimum size=15mm}
    ]

    \node[allnode]    (a2)    at(0,0)  {a2};
    \node[roundnode]    (b2)    at(0,3)  {b2};
    \node[allnode]    (c2)    at(3,0)  {c2};
    \node[roundnode]    (d2)    at(3,3)  {d2};
    \node[allnode]    (e2)    at(6,3)  {e2};
    \path [->]  (b2) edge node[left] {} (a2);
    \path [->]  (c2) edge node[left] {} (b2);
    \path [->]  (d2) edge node[left] {} (c2);
    \path [->]  (b2) edge node[left] {} (d2);
    \path [<->]  (d2) edge node[left] {} (e2);

\end{tikzpicture}
\caption{Example 2: argumentation semantics visualization: preferred (grey) and maximal ideal (dashed border). The grounded extension is $\{\}$. The preferred/maximal ideal argument set is also complete. \{\} is an additional complete set of arguments.}
\label{fig:example2}
\end{figure}

Example 2 highlights the advantage of maximal ideal over grounded semantics for the use case in focus. We start with the following empathic agent scenario:
\begin{itemize}
    \item Acceptability rules persuader: ${(Show \: steak \: ad, *)}$ ($a_2$)
    \item Acceptability rules mitigator: $\{\}$
\end{itemize}
To resolve the acceptability rule inconsistency, the agents exchange the following arguments:
\begin{enumerate}
    \item Mitigator: $\{(a_2, Steak \: ad \: too \: unhealthy)\}$ ($b_2$)
    \item Persuader: $\{(b_2, Steak \: ad \: shows \: quality \: meat)\}$ ($c_2$)
    \item Mitigator: $\{(c_2, User \: prefers \: sweets )\}$ ($d_2$)
    \item Persuader: $\{(d_2, This \: preference \: does \: not \: matter)\}$ ($e_2$)
    \item Mitigator: $\{(e_2, d_2)\}$
    \item Persuader: $\{(c_2, b_2)\}$
\end{enumerate}
From the scenario specification, the following argumentation framework can be constructed:
\begin{align*}
    AF_2 = \langle \{a_2, b_2, c_2, d_2, e_2 \}, \{(b_2, a_2) (b_2, d_2), (c_2, b_2), (d_2, c_2), (e_2, d_2) \} \rangle
\end{align*}
Applying preferred, complete, grounded, and maximal ideal semantics to solve the framework yields the following results:
\begin{itemize}
    \item Preferred: $\{a_2, c_2, e_2\}$
    \item Complete: $\{a_2, c_2, e_2\}$, $\{\}$
    \item Grounded: $\{\}$
    \item Maximal ideal: $\{a_2, c_2, e_2\}$
\end{itemize}
As can be seen, applying grounded semantics removes acceptability rule $a_2$, although this rule is not successfully attacked according to the definition stipulated in our requirements.
In contrast, applying maximal ideal semantics does not remove $a_2$.
Figure~\ref{fig:example2} contains a graphical overview of how the different argumentation semantics solve the argumentation framework of the second example.

\section{Discussion}
\label{discussion}
\subsection{Argumentation and Jason}
\label{argumentation-in-jason}
Argumentation in agent-oriented programming languages has been covered by previous research.
In particular, both Berariu~\cite{berariu2014argumentation} and Panisson et al.~\cite{panisson2014approach} describe the implementation of argumentation capabilities with Jason.
In contrast to these works, we present a service-oriented approach for argumentation support that provides one simple function to solve arguments and thus offers a higher level of abstraction for application developers.
For future work, we suggest considering advanced abstract argumentation approaches like value-based argumentation~\cite{bench2003persuasion} to account for agent preferences and possibilistic argumentation~\cite{alsinet2008logic} to account for uncertainty.
Generally, we consider the approach of writing service interface extensions for Jason as promising to integrate with existing relevant technologies, for example to provide technology bridges to tools and frameworks developed by the machine learning community.

\subsection{Alternative Negotiation Approaches}
\label{alternatives}
We consider the empathic agent approach as fundamentally different from other negotiation approaches (see for an overview: Fatima and Rahwan~\cite{fatima2013negotiation}) in that the empathic agents only optimize their own utility if acceptability rules explicitly permit doing so and strongly consider the utility of other agents in their environment.
I.e., empathic agent application scenarios can be considered at the intersection of mixed-motive and fully cooperative games.
Comprehensive comparisons, in particular with other argumentation-based negotiation approaches, as for example developed by Black and Atkinson ~\cite{black2011choosing}, are certainly relevant future work.
For example, simulations that elaborate on the benefits and limitations of specific empathic agent implementations could be developed and executed in scenarios similar to the one presented in the examples of this paper.

\subsection{Limitations}
\label{limitations}
This paper explains how to implement basic empathic agents with Jason and how to extend Jason to handle scenarios with initially inconsistent beliefs by applying argumentation theory.
However, the implemented agents are merely running examples to illustrate the \emph{empathic agent} concept and do not solve any practical problem.
The following limitations need to be addressed to facilitate real-world applicability:
\begin{itemize}
    \item Although we extended Jason to better support the use case in focus, the current implementation does not provide sufficient abstractions to conveniently implement complex empathic agents. Moreover, our empathic agent implementation does not make use of the full range of BDI-related concepts Jason offers.
    \item While the agents can find consensus in case of simple belief inconsistencies, they do not use scalable methods for handling complex scenarios with large state and action spaces, and are required to explicitly exchange their preferences (utility mappings) with each other, which is not possible in many application scenarios, in which the utility mappings need to be \emph{learned}.
    \item So far our empathic agent only interacts with other empathic agents.
    This simplification helps explain the key concepts with basic running examples and is permissible in some possible real-world scenarios, in particular when designing empathic agents that interact with agents of the same type in closed environments (physically or virtually) that are not accessible to others.
    To interact with agents of other designs, the empathic agent concept needs to be further extended, for example, to be capable of dealing with malevolent agents.
    \item The analysis of the argumentation approach is so far limited to the admissible set-based semantics as introduced in the initial abstract argumentation paper by Dung \cite{dung_acceptability_1995} that considers arguments as propositional atoms. The structure of the arguments is not addressed.
\end{itemize}

\subsection{Towards a Generic Empathic Agent}
\label{towards}
For future research, we suggest further extending Jason--or to work with alternative programming languages and frameworks--to support more powerful abstractions for implementing empathic agents, for example:
\begin{itemize}
    \item Develop engineering-oriented abstractions for empathic agents, possibly following a previously introduced architecture proposal~\cite{kampik_towards_2018}. While further extending Jason can facilitate scientific applications and re-use, providing a minimally invasive library or framework for a general-purpose programming language could help introduce the concept to the software engineering mainstream.
    User interfaces could be developed that facilitate the explainability of the empathic agents' decision processes.
    \item Leverage existing research to better address partial observability and subjectivity. The aforementioned survey by Albrecht and Stone shows that a body of reinforcement learning research from which can be drawn upon exists~\cite{albrecht_autonomous_2018}. Yet, the general problem of scaling intelligent agents in complex and uncertain environments remains an open challenge. It would be interesting to investigate how this problem affects the applicability of our empathic agent.
    \item Design and implement empathic agents that can meaningfully interact with agents of other architectures and ultimately with humans. As a starting point, one could develop an empathic Jason agent that interacts with a \emph{deceptive} Jason agent as introduced by Panisson et al.~\cite{panisson_lies_2018}.
    \item Extend the analysis of a suitable argumentation semantics for the empathic agent by considering advanced approaches, for example value-based argumentation as introduced by Bench-Capon~\cite{bench2003persuasion}, and by identifying a feasible structure the arguments can use internally.
\end{itemize}
\section{Conclusion}
\label{concl}
In this paper, we show how to implement proof-of-concept prototypes of empathic agents with an existing multi-agent development framework.
Also, we devise an extension of the previously established empathic agent concept that employs argumentation theory to account for inconsistent beliefs between agents.
In our analysis of different admissible set-based argumentation semantics, we have determined that maximal ideal semantics are the most feasible for addressing the problem in focus, as they provide both an unambiguous solution of a given argumentation framework and are not overly sceptical (in contrast to grounded semantics) as to which arguments are acceptable.

However, the provided empathic agent implementations do not yet scale to solve real-world problems.
It is important to identify or devise more powerful methods to handle subjectivity and partial observability, and to ultimately enable empathic agents to meaningfully interact with agents of other architecture types.
On the technology side, an abstraction specifically for empathic agents is needed that forms a powerful framework for implementing empathic agents.
We suggest creating a Jason extension that can facilitate the implementation of empathic agents with a framework that is familiar to members of the academic multi-agent systems community, but to also consider providing abstractions for programming frameworks that are popular among industry software engineers and are not necessarily BDI-oriented or even agent-oriented per se.
\subsubsection{Acknowledgements}
We thank the anonymous reviewers for their constructive feedback.
This work was partially supported by the Wallenberg AI, Autonomous Systems and Software Program (WASP) funded by the Knut and Alice Wallenberg Foundation.
%
%
%
\bibliographystyle{splncs04}

\end{document}